# Temperature Driven Transformation of CsPbBr$_3$ Nanoplatelets into Mosaic Nanotiles in Solution through Self-Assembly

Zhiya Dang[†], Balaji Dhanabalan[†ʇ], Andrea Castelli[†d], Rohan Dhall[¥], Karen C. Bustillo[¥], Dorwal Marchelli[‡], Davide Spirito[‡d], Urko Petralanda[†], Javad Shamsi[†§], Liberato Manna[†*], Roman Krahne[‡], and Milena P. Arciniegas[†*]

[†]Nanochemistry Department. [‡]Optoelectronics. Istituto Italiano di Tecnologia, Via Morego 30, 16163 Genova, Italy.

[ʇ]Dipartimento di Chimica e Chimica Industriale, Università degli Studi di Genova, Via Dodecaneso, 31, 16146 Genova, Italy.

[¥]National Center for Electron Microscopy, Molecular Foundry, Lawrence Berkeley National Laboratory, Berkeley, California 94720, United States.

## Abstract

Two-dimensional colloidal halide perovskite nanocrystals are promising materials for light emitting applications. In addition, they can be used as components to create a variety of materials through physical and chemical transformations. Recent studies focused on nanoplatelets that are able to self-assemble and transform on solid substrates. Yet, the mechanism behind the process and the atomic arrangement of their assemblies remain unclear. Here, we present the transformation of self-assembled stacks of CsPbBr$_3$ nanoplatelets in solution, capturing the different stages of the process by keeping the solutions at room temperature and monitoring the nanocrystal morphology over a period of a few months. Using ex-situ transmission electron microscopy and surface analysis, we demonstrate that the transformation mechanism can be understood as oriented attachment, proceeding through the following steps: i) desorption of the ligands from the particles surfaces, causing the merging of nanoplatelet stacks, which first form nanobelts; ii) merging of neighboring nanobelts that form more extended nanoplates; and iii) attachment of nanobelts and nanoplates, which create objects with an atomic structure that resemble a mosaic made of broken nanotiles. We reveal that the starting nanoplatelets merge seamlessly and defect-free on an atomic scale in small and thin nanobelts. However, aged nanobelts and nanoplates, which are mainly stabilized by amine/ammonium ions, link through a bilayer of CsBr. In this case, the atomic columns of neighboring perovskite lattices shift by a half-unit-cell, forming Ruddlesden-Popper planar faults. We also show that the transformation is temperature-driven and that it can be accelerated to the point that it can take place within tens of minutes in solution and in spin-coated films. In-situ monitoring of the nanocrystal photoluminescence spectra, combined with transmission electron microscopy analysis, indicates that the same mechanism is at work for the accelerated transformation at high temperature. Understanding this process gives crucial information for the design and fabrication of perovskite materials, where control over the type and density of defects is desired, stimulating the development of perovskite nanocrystal structures with tailored electronic properties.

Metal halide perovskites offer fascinating chemical and structural versatility, coupled with excellent optical and electronic properties, enabling them to be applied to many different optoelectronic devices, such as solar cells, light emitting diodes, lasers, and photodetectors.[1-4] In their nanocrystal form, the emission wavelength from metal halide perovskites can be easily tuned over a broad range of the visible spectrum, by changing the particle size, dimensionality, or cation and anion composition.[5, 6] Thanks to recent progresses in the development of new synthesis methods, it is now possible to prepare ligand-passivated nanocrystals through different approaches (e.g. by the heating-up method, or the ligand-assisted re-precipitation technique).[5, 7-11] However, due to the high dynamicity of the molecules stabilizing the particle's surfaces,[12] as well as the ionic nature of perovskite nanocrystals and the ease at which ion migration can occur inside them, they are found to be intrinsically unstable. Nevertheless, the surface chemistry of metal halide perovskite nanocrystals can be exploited to promote and control physical-chemical transformations. The nature of the surface is particularly relevant in the case of two-dimensional nanocrystals, such as nanoplatelets (NPLs), nanosheets and nanodisks, which are highly anisotropic particles that can have atomically precise thicknesses and can exhibit strong quantum confinement effects.[13-17] Such structures are prone to self-assemble and then undergo oriented attachment, a process by which the nanocrystals achieve a lattice match, and eventually connect to each other and build larger single objects under the cooperative effects of short- and long-range interactions.[13, 15, 18-22] In this process, adjacent nanocrystals with identical crystal facets that face one another undergo continuous rotation and rearrange their atoms through the formation of a neck in the region of contact, until they become a single structure.[23, 24]

Perovskite nanocrystals have the ability to undergo shape- and phase-transformation by self-assembly, which has been recently exploited to fabricate nanoplates and nanosheets from $CsPbBr_3$ nanocube superlattices under an external pressure,[25] and to produce nanosheets by applying solvothermal conditions to $CsPbX_3$ (with X = Cl, Br, and I) nanorods that interact side-by-side.[26] In the building steps of such structures, the oriented attachment between neighboring nanocrystals plays a key role, as it has been also demonstrated for other types of nanocrystals.[27-29] Both the transformation and the assembly geometry are affected by the nanocrystal concentration in solution, since the concentration affects the distance between the individual particles. Increasing the particle concentration enhances strong ligand-ligand interactions, leading to long-range ordered structures; a diluted suspension instead favors ligand destabilization and therefore extended sheets can form via crystallographic oriented attachment.[18] Other driving forces, such as laser irradiation, have been reported for the transformation of self-assembled NPLs into nanobelts[10] or bulk structures,[30] which improved the stability of the optoelectronic devices, and could be used to modify the emission wavelength, exploiting the particles for color patterns. Mostly, such transformations were investigated in the solid state, which limits the nanocrystal ability to move freely, and hinders ligand mobility and interactions. Instead, a bulk liquid environment acts as a media that facilitates the overall transformation. However, a detailed investigation of the transformation of two-dimensional perovskite nanocrystals via self-assembly in solution is still missing.

In this work, we study the spontaneous and heat-induced transformation of self-assembled $CsPbBr_3$ NPLs into larger nanostructures such as nanobelts, nanoplates and nanotiles in bulk solution. For a detailed study of the mechanism, the evolution in the morphology and atomic structure of the nanocrystals was examined over time under ambient conditions using ex situ transmission electron microscopy (TEM) analysis. Such structural evolution is associated with changes in the photoluminescence (PL) of the objects that are produced at the different stages of the transformation. Initially, the NPL stacks present in the freshly prepared solution merge either face-to-face or side-to-side, giving rise to intermediate products such as nanobelts, which were widely observed in aged solutions and which caused a red-shift in the PL emission. These structures are found together with $Cs_4PbBr_6$ hexagonal-shaped nanocrystals, which typically are formed by perovskite $CsPbBr_3$ nanocrystals reacting with an excess of amines in solution. Therefore, the emergence of $Cs_4PbBr_6$ nanocrystals in the solution evidences the presence of desorbed ligand molecules from the surface of the $CsPbBr_3$ NPLs, triggering their transformation. At later stages, the self-assembly of the intermediate products occurs, and the assembled structures merge into larger nanoplates. Eventually, the aged components from the different stages of the transformation in the solution attach to each other in a mosaic manner and create even larger objects, such as nanotiles, which emit at the same wavelength as bulk $CsPbBr_3$. Each stage is facilitated by the oriented attachment of adjacent components through the rearrangement of atoms at the connecting facets. We found that this atomic rearrangement enables the formation of defect-free boundaries in the early formation of nanobelts and nanoplates, producing a continuous perovskite atomic lattice. In contrast, the atomic structure at the boundaries of neighboring domains resulting from aged objects is often imperfect, forming mosaic-like nanotiles in solutions that were stored for more than a month. The mosaic patterns arise from the presence of CsBr bilayers that are located at the interface, and the lattice mismatch is evident from the atomic columns, which are shifted by half a unit, spontaneously generating local Ruddlesden-Popper planar faults. These are uncommon defects in halide perovskites (so far, they have been observed after a post-synthesis treatment and in mixed halide nanocrystals).[31-34] This analysis of the assembly and transformation of $CsPbBr_3$ NPLs in solution brings a key mechanistic understanding into the evolution of such low-dimensional nanocrystals that can be exploited for the active design of desired objects. Furthermore, perovskite NPLs as 'transformer materials' are promising candidates for investigating defects in halide perovskite crystals and assessing their impact on the electronic properties of the resulting structures. Towards practical implementation, we demonstrate that the NPL assembly and transformation process can be significantly accelerated in solution and thin films by controlled heating. Spin-coated NPL films were transformed within less than 30 min by heating to 110 °C, showing the same intermediate photoluminescence spectra and the same final nanocrystal morphologies. The implementation of the transformation into device fabrication procedures in short time-scales unlocks the technological potential of these solution-based processes.

**Results and Discussion**

**Starting nanocrystals.** We used $CsPbBr_3$ NPLs that were synthesized at a relatively low temperature (60 °C) in the presence of octadecene, oleic acid and oleylamine (see the Methods section for further details). The as-synthesized NPLs have a length of ca. 21 nm, a width of ca. 8 nm, and a thickness of 3 nm. The NPLs in freshly prepared solutions spontaneously form stacks, in which they have a particle-particle distance of ca. 2 nm, as was determined by TEM analysis (see Figure 1a and Figure S1 of the Supporting Information (SI)). Here, the initial concentration of NPLs in toluene based on the Pb content was 9 μM, as determined by inductively coupled plasma mass spectroscopy (ICP-MS). The selected area electron diffraction (SAED) pattern that was acquired on the initial NPLs confirms their $CsPbBr_3$ structure (Figure S2a, c). The fresh NPLs exhibit blue emission under ultraviolet (UV) light excitation, and their emission peak is centered at 460 nm, which matches previous reports for similar structures.[10] The vials containing the nanocrystal suspensions used for studying their transformation over time were covered with Al foil and placed on a closed cabinet to avoid potential effects from light exposure. The vials were kept at room temperature with a ca. 50% of relative humidity, without shaking.

**Transformation route.** To monitor the evolution of the as-synthesized NPLs, we collected aliquots from the nanocrystal suspensions at different points in time and investigated their morphology with TEM. Figure 1 displays a collection of TEM images from fresh (Figure 1a) and aged solutions, recorded after 1-week, 1-month, and 2-months (Figure 1b-d). TEM images were acquired from nanocrystal solutions that had been deposited on carbon-coated Cu grids by drop casting and were left to dry, and the morphology analysis was conducted on several regions of the TEM grids. Images recorded after storage of 1 week show well-defined structures such as nanobelts (Figure 1b). The width of the nanobelts (ca. 21 nm) is the same as the length of the initial NPLs, which suggests that they are formed as a result of a merging process of the as-synthesized $CsPbBr_3$ NPL stacks. Images recorded at longer storage times (1 month) document that these structures then evolve into larger nanoplates, and eventually transform into extended nanotiles when they are left to age for 2 months (Figure 1c-d). These morphological changes are supported by a statistical nanocrystal size analysis (see Table S1 and Figure S3) on images that were collected from different regions. Overview TEM images are provided in Figures S4-S7. The size of the observed structures, in terms of projected area in TEM, increases dramatically, from an average of 66 $nm^2$ (NPLs) to ca. 35,000 $nm^2$ (nanotiles) after two months of aging, while their number is reduced from ca. 3,500 NPLs/$\mu m^2$ to ca. 8 nanotiles/$\mu m^2$, respectively. In the initial stage (after 1 week), the resulting nanobelts have different lengths because the merging stacks contain a different number of NPLs (Figure S8). After one month, we observed fewer nanobelts and a larger number of nanoplates, as highlighted by a white dashed line in Figure 1c. When the solution was left to age for more than a month, larger and thicker structures (the nanotiles) were formed and became the dominant population (Figure 1d). Throughout these stages, the $CsPbBr_3$ crystal structure was preserved (orthorhombic phase, with ICSD number 97851), as was confirmed by the collected SAED patterns (Figure S2b, c) and by Energy Dispersive X-ray Spectroscopy (EDS) via Scanning TEM (STEM) (Figure S9 and Table S2).

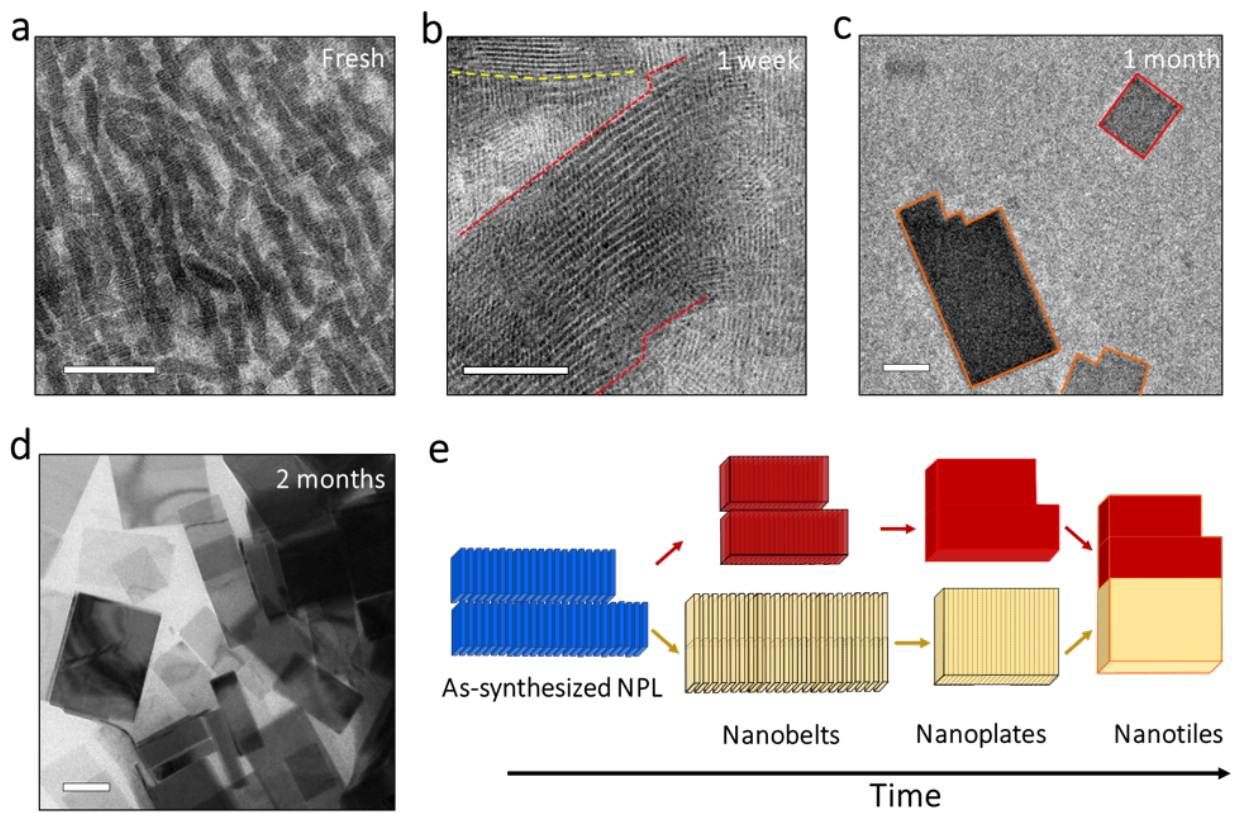

**Figure 1. A collection of TEM images showing the structural transformation of CsPbBr$_3$ NPLs.** (a-d) Representative TEM images of fresh NPL stacks (a), which evolve into nanobelts (b) after a short time (~ 1 week). Longer aging (~ 1 month) leads to different sized nanoplates in the solution, along with a few nanotiles (which are framed by a red line in (c)). These nanotiles become the dominant products in solutions that have been aged for 2 months (d). Scale bars: 100 nm. (e) A scheme of the transformation process of as-synthesized CsPbBr$_3$ NPL stacks in solution over time: Stage I, the formation of nanobelts via a face-to-face (in red) and/or side-to-side (in yellow) merging of NPL stacks; Stage II, the assembly of nanobelts that form nanoplates; Stage III, the attachment of nanobelts and nanoplates that create mosaic-like nanotiles.

The transformation of NPL stacks into nanotiles over time can be broken down into three stages as illustrated in Figure 1e: (Stage I) the NPL stacks move freely in solution and neighboring NPLs within a stack, which are facing one another with identical (001) facets (considering the unit cell of orthorhombic phase with ICSD number 97851, Figure S2), merge face-to-face via crystallographic oriented attachment; a similar process is followed by stacks in close proximity that face each other with identical (110) facets and fuse side-by-side, resulting in the formation of nanobelts; (Stage II) the nanobelts self-assemble and merge into nanoplates; and (Stage III) nanobelts and nanoplates attach in a mosaic manner, following a contact through identical facets, creating large nanotiles.

To correlate the observed structural evolution of the NPL stacks with their optical properties in solution, we collected absorption and PL spectra at different points in time during the aging process (Figures S10 and S11). The initial blue emitting solution has an emission peak at 460 nm (Figure S10), which is red-shifted to 463 nm after 1 week. This red-shift stems from the aggregation of NPL stacks (as seen in Figure 1b), and can be related to changes in their dielectric environment.[35,36] A new broad emission peak with a relatively low intensity arises at 520 nm (Figure S11), and is associated with the formation of a few larger objects such as nanobelts in the solution. The blue PL peak of solutions that are left to age for longer time further red-shifts (to 465 nm after two weeks and to 470 nm after one month), and decreases in intensity (Figure S10). In parallel, the intensity of the green emission peak observed at 520 nm increases, which is in line with the increased formation of larger structures such as nanobelts. The green emission peak further red-shifts to 525 nm after 2 months. Since this wavelength corresponds to the emission of bulk $CsPbBr_3$, it indicates the formation of structures that are larger than those in the quantum confinement regime. The emission then remains stable at 525 nm (monitored for up to 6 months of aging).

**Surface chemistry.** Importantly, in the TEM images of 1 week-aged solutions we observed hexagonal structures with an edge length of around 400 nm. These structures occurred randomly together with the initial NPL stacks (Figure S8). Figure 2a shows a High Angle Annular Dark Field-STEM (HAADF-STEM) image of two of these hexagonal structures. Elemental analysis via STEM-EDS shows that such structures have a Cs:Pb:Br ratio of 4:1:6, indicating that they are $Cs_4PbBr_6$, so-called zero-dimensional nanocrystals (see Figure S12 and Table S3 for more details). Such $Cs_4PbBr_6$ nanocrystals have been reported to form by a post-synthesis transformation of perovskite nanocrystals with an excess of amines.[37, 38] This transformation takes place because the excess amines extract PbBr$_2$ from the perovskite nanocrystals to form PbBr$_2$ complexes. Therefore, the emergence of Cs$_4$PbBr$_6$ structures indicates that there is an excess of amine species in the CsPbBr$_3$ NPL solutions after a relatively short time of aging, which could be caused by ligands that were released from the particles' surfaces. Among these ligands, oleylamine can be desorbed and react with the perovskite NPLs, converting the particles into Cs$_4$PbBr$_6$ (see the scheme in Figure 2a). The release of oleylamine, in the form of oleylammonium ion, typically involves the concomitant release of a counter ion, in this case Br$^-$. The removal of Br$^-$ facilitates the accommodation of Cs$^+$ ions at the surface of the particles.[39] The ligand desorption most likely occurs when NPL stacks merge into nanobelts, a process that involves a reduction of the surface area. In this scenario, the majority of the ligands that coat the NPLs' surfaces (i.e. oleate, oleylammonium, etc.)[40, 41] are released into the solution, as is illustrated in Figure 2b. Such a desorption process is possible because of the highly mobile nature of ligands on the perovskite nanocrystals,[12] which, in this case, leads to merging of the NPLs. In order to qualitatively assess the role of the ligands in the transformation process, we investigated the surface of the crystals in fresh and aged (4 month-old) NPL solutions via Fourier Transform Infrared (FTIR) Spectroscopy. To maximize the footprint of bound ligands, we performed the FTIR comparative analysis on dried samples that were prepared from highly concentrated suspensions, by drop casting an aliquot onto the surface of the ATR (attenuated total reflectance) crystal and allowing full solvent evaporation in open air (see Methods). After the NPLs synthesis, the suspension was washed once to remove mainly unreacted precursors and the excess of ligands in the solution, without reducing the surface capping of the particles or damaging the NPL surfaces. We ensured that the samples completely covered the ATR-FTIR spot, thus guaranteeing that the signal comes from a large number of nanocrystals in direct contact with the ATR crystal surface.

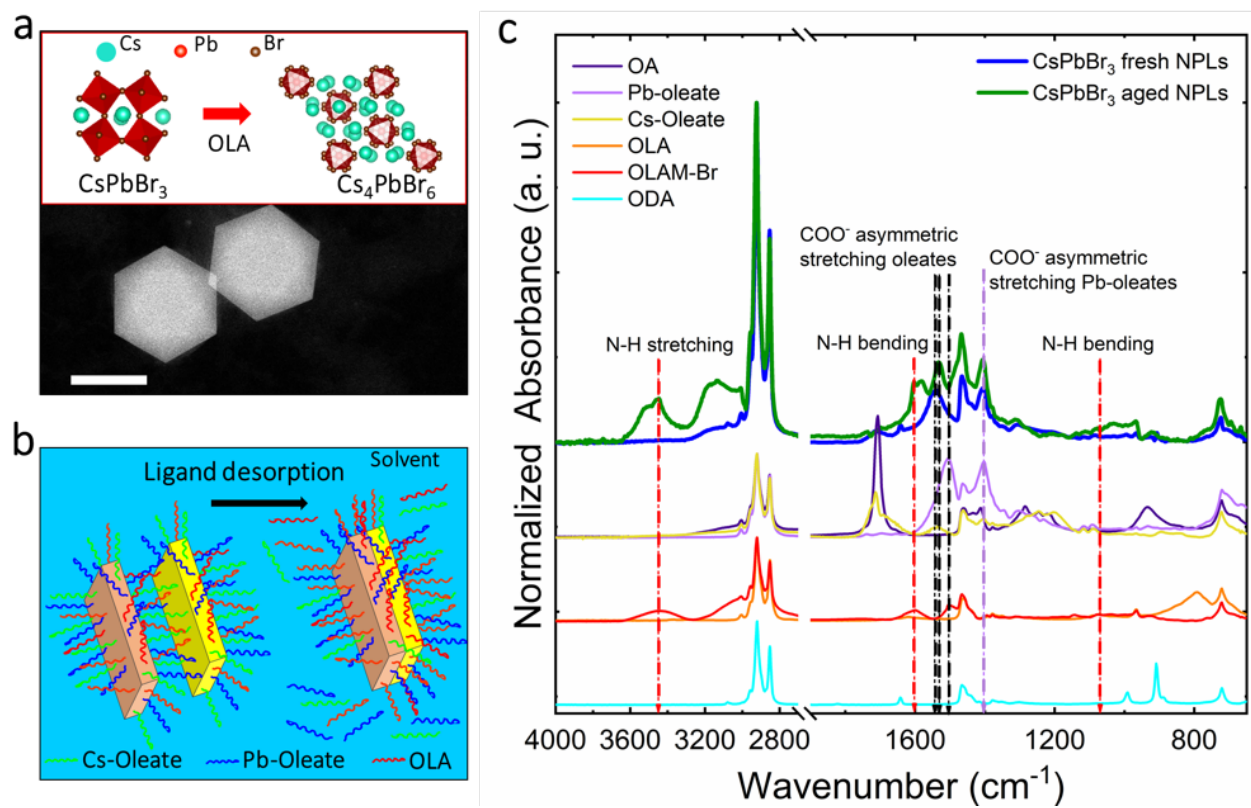

**Figure 2. Surface chemistry evolution during the NPL transformation.** (a) A HAADF-STEM image of hexagonal-shaped crystals from the 1-week aged solution. According to EDS compositional analysis, these crystals have a $Cs_4PbBr_6$ structure. Scale bar: 500 nm. The illustration displays the transformation of the orthorhombic $CsPbBr_3$ phase into a hexagonal $Cs_4PbBr_6$ phase when amine species are present in the solution. Both crystal models show the unit cell oriented in zone axis [001]. (b) A scheme illustrating the ligand desorption during the merging of two NPLs. (c) FTIR absorption spectra of the dried nanocrystals from fresh (blue line) and 4 month-old (green line) NPL solutions compared to the spectra recorded from chemicals employed in their respective syntheses: oleic acid (OA), Pb- and Cs-oleate, oleylamine (OLA), oleylammonium bromide (OLAM-Br), and octadecene (ODA). The vertical lines indicate the characteristic vibrational peaks of the nanostructures in the aged NPL nanocrystals (nanotiles), which show changes that are associated with an increase in the amount of oleylamine and its compounds, as well as a reduction in the contribution from Cs-oleate.

For comparison, we collected the FTIR spectra from the pure chemicals that were used in the NPL synthesis, namely octadecene (ODA), oleic acid (OA), Cs-oleate, Pb-oleate, oleylamine (OLA), and oleylammonium bromide (OLAM-Br as a product of the reaction of oleylamine with $PbBr_2$). These chemicals mimic the vibrations of the possible ligands that coat the nanocrystals. The FTIR spectra are displayed in Figure 2c, and we identified two regions of interest: from 3750 cm$^{-1}$ to 2750 cm$^{-1}$, which contains the various N-H and C-H stretching modes of all the ligands; and from 1800 cm$^{-1}$ to 600 cm$^{-1}$, which comprises the vibrational markers of the ligands (see the detailed assignment of the absorbance peaks in Table S4).[42-44] The most intense peak, related to CH$_2$ asymmetric stretching, is centered at 2925 cm$^{-1}$, and it was used to normalize the absorbance intensities in the spectra. Compared to NPLs, the FTIR spectrum of nanotiles exhibits three new vibrational features that are indicated by vertical red dotted lines: a broad peak at ca. 3500 cm$^{-1}$, a double peak at 1600-1580 cm$^{-1}$, and another broad peak around 1040 cm$^{-1}$. These features are all ascribed to the N-H vibrational modes of the primary amines and ammoniums; in this case, OLAM-Br (see FTIR spectrum of pure OLAM-Br vs OLA). The relatively large width of these absorbance peaks indicates that OLAM-Br is bound to the surface of the nanotiles. In addition, these nanocrystals show more intense absorbance peaks in the region from 1550 cm$^{-1}$ to 1350 cm$^{-1}$, with a peak at ca. 1405 cm$^{-1}$ (indicated with a violet vertical line) that corresponds to the COO$^-$ vibrational marker of Pb-oleate, as indicated by its strong intensity in the FTIR spectrum of the pure Pb-oleate. There is also an absorbance peak centered at 1530 cm$^{-1}$, which shifted 10 cm$^{-1}$ to lower wavenumbers with respect to the one that was observed for fresh NPLs (at 1540 cm$^{-1}$). Both peaks denote a COO$^-$ stretching mode of oleates (see the vertical black lines in the spectra). Since the vibrational marker for Cs-oleate is at 1540 cm$^{-1}$, and that of Pb-oleate is at 1510 cm$^{-1}$, the observed peak shift in the spectra of the nanotiles, along with its increased intensity, indicates that Pb-oleate contributes more to the absorbance than Cs-oleate in the case of the aged nanocrystals. The signal from C=O stretching at 1710 cm$^{-1}$ and =C-H wagging at 900 cm$^{-1}$ that are vibrational markers for OA and ODA, respectively, is relatively weak, which indicates that these compounds are present in minor traces, thus their role is mostly that of solvents. We conclude that the aged structures contain relatively more amine/ammonium and less Cs-oleate on their surfaces than the as-synthesized NPLs, which confirms that, after the NPL stacks merge, a different passivation mechanism is activated. One possibility is that Cs ions are replaced by oleylammonium ions, which then make strong bonds with the surrounding Br$^-$, as has been previously demonstrated for similar systems.[12, 45] Therefore, the spontaneous transformation of NPLs into nanotiles over time leads not only to a transformation in nanocrystal shape, but also to a more stable ligand passivation of the surface.

**Atomic arrangements in the transformed nanostructures.** To closely examine the transformed structures at different stages, we used HAADF-STEM and high resolution (HR) TEM. Figure 3 displays HAADF-STEM images of representative nanobelts collected from the 1-month aged solution. In these images, the bright spots correspond to the Pb-Br atomic columns (highlighted with red dots in Figure 3b) due to their higher average atomic number as compared to the Cs-Br and Br ones. The nanobelts manifest a continuous atomic structure (Figure 3a). These structures stem from the merging of the initial NPL stacks, and their defect-free atomic structure demonstrates the perfect crystallographic lattice resulting from the rearrangement of atoms during the oriented attachment of neighboring NPLs and consecutive NPL stacks. The continuous atomic lattice inherent to the nanobelts was also observed when individual NPLs joined an already existing nanobelt, as shown in Figure 3b (which is a magnified view of the white framed region in the inset). The white dotted lines follow the edges of the merged structures, while the black dotted line highlights their boundary. More examples of perfectly attached structures are depicted in Figure S13. We conclude that the assembly and merging of NPLs in Stage I and Stage II of the transformation process is typically characterized by a defect-free atomic binding at the interface between the merging components. This is illustrated by the atomic model in Figure 3c, in which atoms are aligned at the boundaries between the NPLs and the nanobelts and form a perfect perovskite lattice.

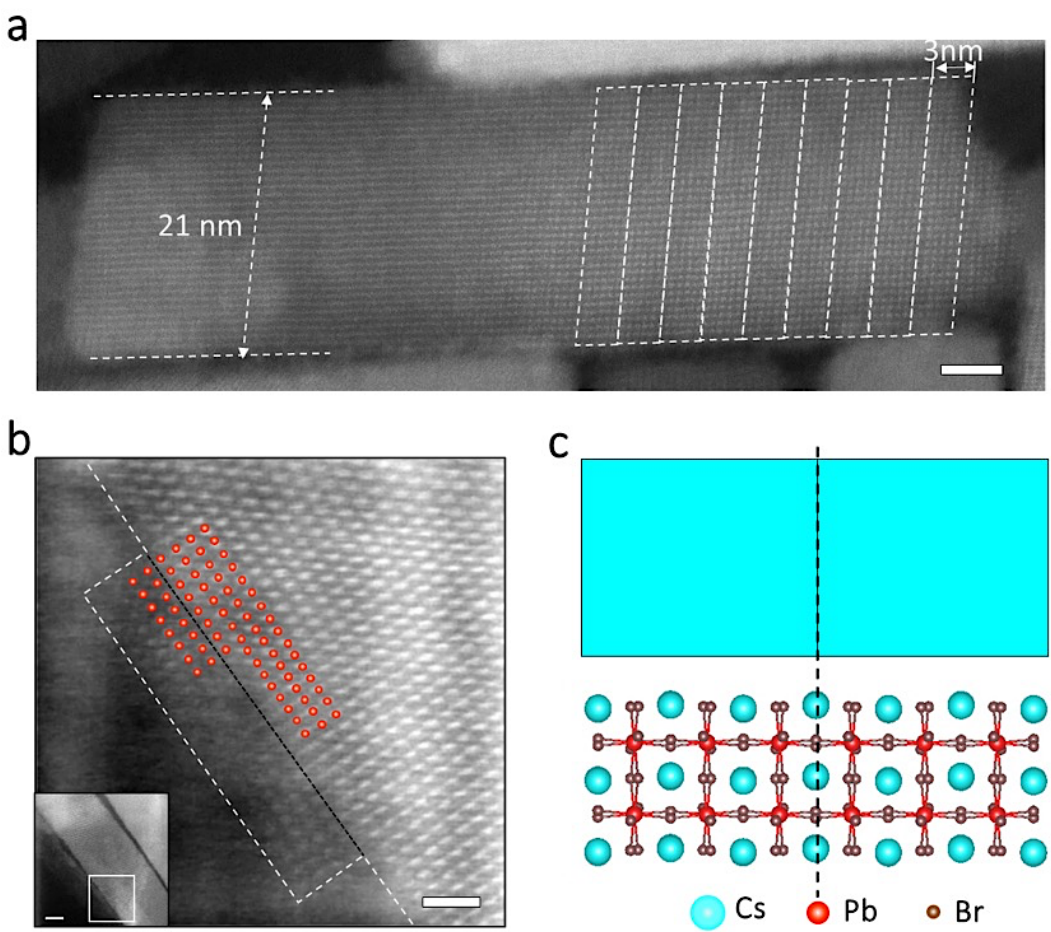

**Figure 3. Perfectly attached atoms at the boundary of merged NPL stacks and assembled nanobelts.** (a) A high resolution HAADF-STEM image showing the defect-free atomic structure of a representative nanobelt formed by merged NPL stacks (framed in white dashed lines) that preserves their original width of ca. 21 nm. Scale bar: 5 nm. (b) A NPL that is bound to the long nanobelt shown in the inset (white framed region). The black dotted line indicates the boundary between the two different components and the white line frames a section of the resulting object. The red dots highlight the Pb-Br atomic columns. Scale bars: 2 nm. (c) An atomic model showing the observed perfect oriented attachment among two NPLs sketched on the top of the model in light blue.

Interestingly, the situation is very different for the attachment among larger structures, i.e. between large nanoplates or nanobelts and nanoplates. For example, in the nanotiles that were formed in the 2 month-old solutions, internal boundaries are clearly visible and create mosaic-like structures, as is highlighted in Figure 4a. To analyze this attachment in more detail, a set of five different components (nanobelts and nanoplates) that attached and formed a region of a nanotile is shown in Figure 4b; their boundaries are highlighted with yellow arrows.

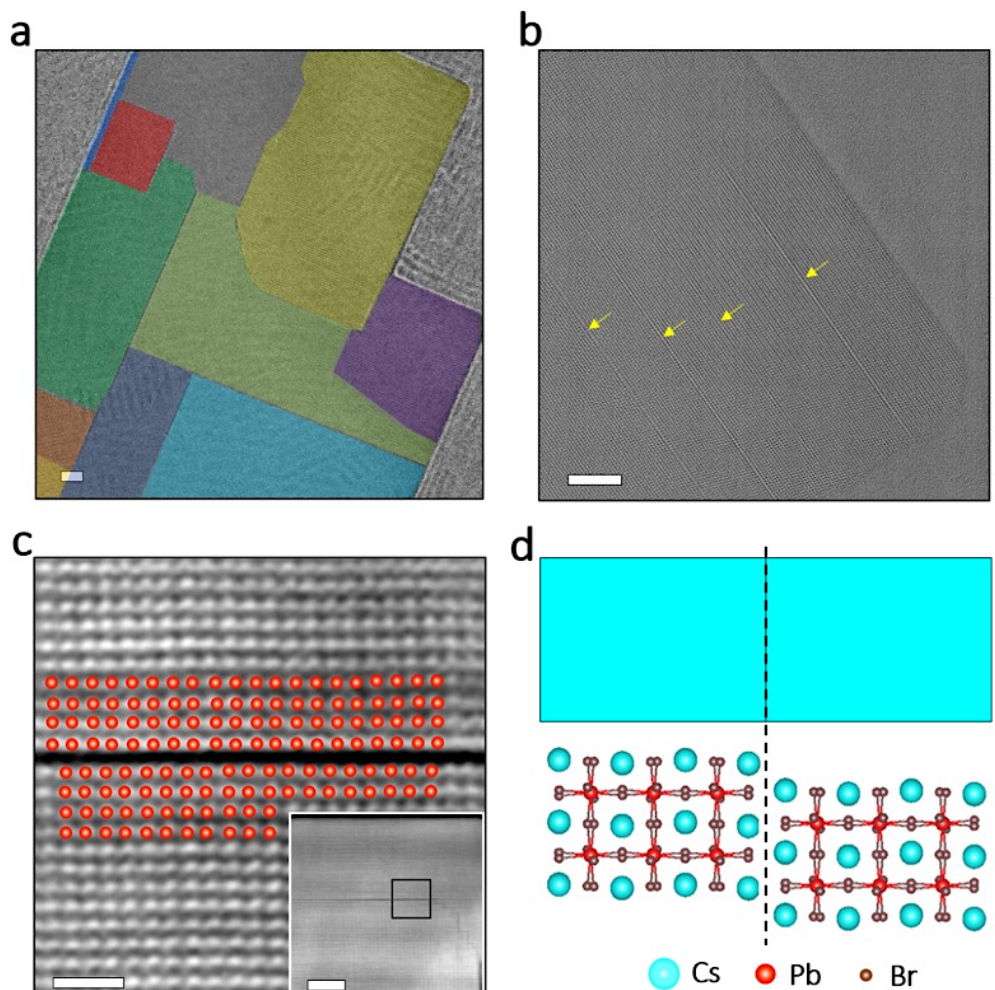

**Figure 4. Ruddlesden-Popper stacking faults in nanotiles.** (a-b) High-resolution TEM images of representative mosaic-like nanotiles. The different components from which the nanotile in (a) was made are highlighted with various (false) colors and the boundaries of the domains within the nanotile shown in (b) are indicated with yellow arrows. Scale bars: 10 nm. (c) A magnified HAADF-STEM image of the boundary displayed in the inset (region framed in black) showing two neighboring domains within a nanotile with their Pb-Br atomic columns (highlighted with red dots) shifted by half of a unit cell. Scale bars: 2 nm; (inset): 10 nm. (d) An atomic model that illustrates the imperfect attachment of two components (light blue rectangle on the top) that form a nanotile.

The view of the atomic structure at these boundaries in Figure 4c demonstrates that adjacent domains have an imperfect attachment. That is, compared to a continous perovskite lattice (as is shown in Figure 3b-c), a Pb-Br plane is missing at the boundary. Instead, the two merging objects are terminated with Cs-Br planes and, upon their attachment, a CsBr bilayer is formed at these boundaries. As a result, there is a spacing of ca. 8.7 Å between the outermost Pb-Br planes of two merging objects at their interface (see Figure S14c, d). Additionally, the perovskite lattices of the neighouring domains are shifted by half a unit cell, as is shown in the magnified HAADF-STEM view in Figure 4c (in which the Pb-Br atomic columns are highlighted with red dots). This shift of half a unit cell is also illustrated in the atomic model in Figure 4d. More examples of objects containing imperfect attachments are shown in Figure S14. The NPL transformation also leads to the formation of other types of defects that are less abundant, such as dislocations and grain boundaries, as is shown in Figure S15. The shift of the atomic column, which is mainly observed in components larger than 50 nm, corresponds to Ruddlesden-Popper (R-P) planar faults, which are generally reported for oxide perovskites[46-50] but have been rarely observed in lead halide perovskite nanocrystals.[31, 32] The presence of planar defects in the resulting nanotiles, as well as their imperfect attachment, can be explained in part by a reduction in the free motion of the larger objects (nanobelts and nanoplates) during later stages of the transformation, which hinders their ability to collide and rotate with respect to close structures via Brownian forces. Thus, the crystallographic attachment occurs among slightly misaligned structures so they are not parallel, a condition needed for their perfect attachment. On the other hand, the FTIR analysis combined with the HRTEM observations points toward CsBr-terminated nanobelts, in which some Cs ions on the surface have been replaced by oleylammonium ions, which results in a more stable surface passivation. As a consequence, their attachment might not involve the removal of Cs atoms, as it occurs in the merging of the NPLs. Hence, they directly link through a CsBr bilayer, accommodating their atoms dislocated by half a unit cell to minimize the energy of the system, which leads to the formation of the unusual R-P planar faults, in a similar way as in oxide perovskites.[48, 49]

To gain a deeper understanding of the formation of the atomic attachment of objects at different stages of the transformation process, we performed density functional theory (DFT) analysis (see details in the Methods section). We calculated the energy of two domains made of a few unit cells that had perfect lattice attachment and compared it to the energy of two domains with a lattice containing a R-P planar fault. The computational analysis that is displayed in Table S5 shows that the formation of a R-P planar fault requires more energy than a perfect lattice ordering, and that the difference is smaller for the merging of two large domains than for two small domains (see Table S5). Thus, the formation of R-P faults is more likely to occur in the merging of the larger objects at the later stages of the transformation process.

**Temperature dependence of the transformation.** The dynamics of the surface passivation that we discussed earlier have important technological implications. They point out that the ligand desorption, and the formation of interfaces based on R-P faults, which are at the base of the transformation process, should strongly depend on temperature. In particular, ligand desorption should increase with increasing temperature, and decrease when the temperature is reduced. As a consequence, heating or cooling can be used to accelerate the transformation process or to preserve the properties of the original NPL solution. We tested heating of the NPL solutions at 50 °C for 1 hour and at 110 °C for 30 min, and cooling at -4 °C for one month. The heated solutions were green emitting after the treatment (Figure S16), and in particular for the solution heated at 110°C we observed the full transformation to nanoplates and nanotiles (Figure S17), which confirmed that the process can be significantly accelerated at higher temperature. Interestingly, we observed a rich presence of R-P faults in the nanotiles, with lateral size from 40 to 200 nm, in the solution heated at 110 °C, which proves that the R-P faults are stable under these conditions (Figure S18). Milder temperature treatment such as heating at 50 °C for one hour promoted the formation of thicker NPLs (Figure S17). On the other hand, the solution stored for 1 month at -4 °C preserved its blue emission and the NPL morphology (Figure S19).

Processing nanocrystal solutions for the fabrication of thin films is one of the major application routes in their use for light emission and energy harvesting. We therefore fabricated thin nanocrystal films from the NPL solutions by spin-coating (see details in Methods) and monitored their photoluminescence (PL) at different temperature over time. At room temperature we observed a full shift of the PL signal to green emission within 3 weeks, indicating a complete transformation process, and at 110 °C this transformation occurred within 40 min., as demonstrated in Figure 5.

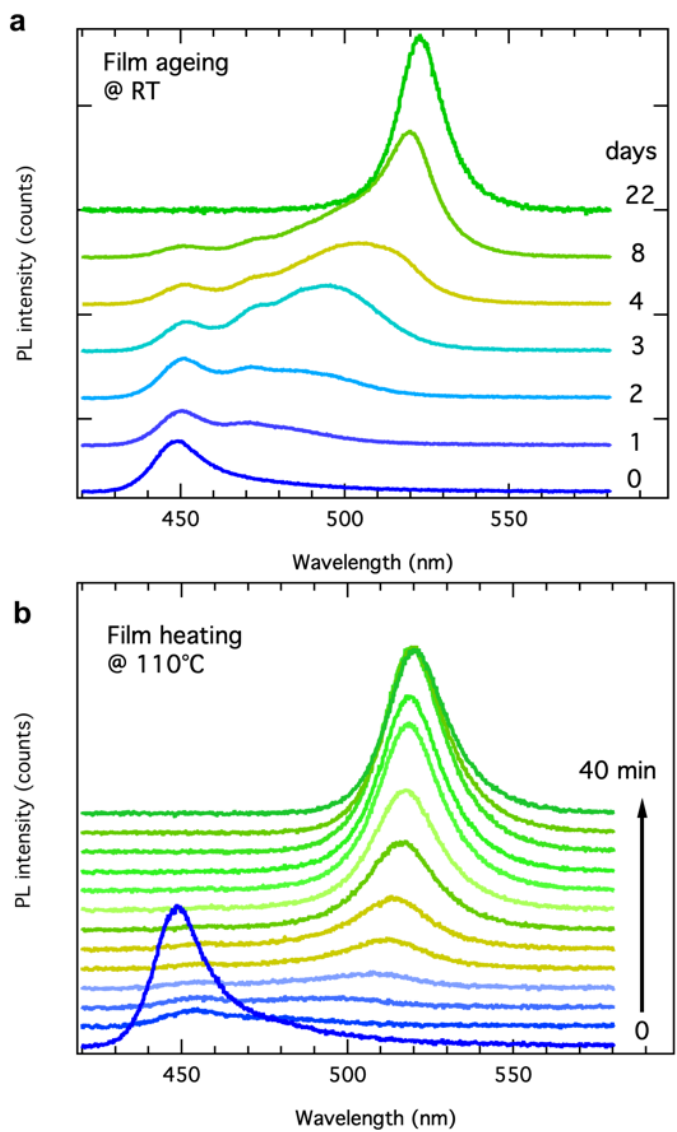

**Figure 5. NPL transformations in spin-coated films.** (a) PL spectra of a film fabricated from NPL solution stored at room temperature recorded over a period of 22 days. (b) In-situ recorded PL spectra over a time period of 40 min of a film heated at 110 °C.

The trend in this transformation is similar to that observed from the aliquots extracted from the solution displayed in Figure S11. Over time, the intensity of the blue emission, initially at 460 nm, decreases, and a PL peak in the green at 515 nm builds up that gradually red-shifts with time until it stabilized at around 530 nm. The transformation in the films also modified their electrical properties. Initial pristine films did not show any reproducible conductance, whereas the heat-transformed films manifested stable photoconductivity with a similar performance as it was obtained by UV-light induced transformation (Figure S20).[10]

**Conclusions**

We investigated the transformation of CsPbBr$_3$ NPLs, starting from their self-assembled stacks in solution via a combined TEM and surface analysis, and elucidated the key mechanisms in the merging of smaller high-surface area nanocrystals to form larger, and bulkier structures. Freshly prepared NPLs merge into single crystalline nanobelt structures by face-to-face or side-to-side oriented attachment of matching facets. The aged nanobelts and nanoplates develop a less rich CsBr surface where oleylammonium ions have replaced Cs$^+$ ions and provide a more stable surface passivation. Oriented attachment of these larger objects with CsBr surfaces leads to the formation of R-P stacking faults, which confers to the resulting nanotiles a mosaic-like architecture with atomic stacking faults. The overall mechanism, based on ligand desorption, particle motion and oriented attachment is temperature-dependent and thus, structures with abundance of R-P faults can be formed in an accelerated way. Such information is crucial for the development of stable perovskite materials for optoelectronics, where the band gap (and thereby emission wavelength), particle shape, and the type and density of defects should be controlled. Furthermore, the transformations routes that we investigated can stimulate the design of novel, and possibly switchable, perovskite materials.

**Methods**

**Synthesis of CsPbBr₃ nanoplatelets.** The nanocrystals were prepared by dissolving 0.145 g of PbBr$_2$ in 4 ml of octadecene, 1 ml of oleic acid, and 1 ml of oleylamine. The mixture was magnetically stirred for 20 min at 100 °C. Next, 0.325 g of Cs$_2$CO$_3$ were dissolved in 5 ml of oleic acid at 100 °C for ca. 15 min. Once both mixtures reached room temperature, 0.5 ml of the Cs-oleate solution were added to 6 ml of the PbBr$_2$ solution. The resulting solution was heated at 60 °C for 30 min under stirring. The mixture was then cooled down by immersing the vial in ice water for 5 min. 2 ml of toluene were added to the product, and the solution was washed once by centrifugation at 3500 rpm for 10 min and the nanocrystals re-dispersed in toluene via sonication for 10 min. All the synthesized suspensions of nanocrystals were stored at room temperature and at ca. 50 % of relative humidity. Elemental analysis of the fresh NPLs was performed via inductively coupled plasma mass spectroscopy by using a Thermo Fisher iCAP 600 instrument. The samples were digested overnight in an HCl solution and diluted in deionized water. All the suspensions were filtered before analysis by using PTFE filters.

**Structural and optical characterization.** TEM analysis was conducted by drop casting suspensions of nanocrystals on carbon-coated copper grids. The suspensions were prepared with different aging times (24 h, 48 h, 72 h, 1 week, 2 weeks, 1 month and two months), including one suspension with fresh particles. An initial shape assessment of the as-synthesized NPLs was performed by using a JEOL JEM-1400 operating at 120 kV. High resolution TEM images were acquired on a JEOL 3010 microscope operating at 300 kV, a FEI ThemIS 60-300 STEM/TEM microscope operating at 300 kV and a JEOL 2200FS microscope. The JEOL 3010 microscope and the FEI were used to acquire the Selected Area Electron Diffraction (SAED) of the structures. The JEOL 2200FS was equipped with a Schottky emitter operating at 200kV, a CEOS spherical aberration corrector of the objective lens, and an in-column energy filter (Omega-type). The JEOL 2200FS was used to collect High-Angle Annular Dark Field Scanning TEM (HAADF-STEM) images of the structures at different times, and the FEI was used to perform an Energy Dispersed X-ray Spectroscopy (EDS) analysis. The Br K-edge and Cs and Pb L-edges were used for all of the TEM-EDS maps of the structures.

The surface of the nanocrystals was characterized using a Fourier transform infrared spectrometer (Equinox 70 FT-IR, Bruker) coupled with an attenuated total reflectance (ATR) accessory (MIRacle ATR, PIKE Technologies). The analysis was performed on samples from highly concentrated suspensions of the as-synthesized NPLs in their fresh and 4 months aged condition. After strongly shaking the nanocrystal suspensions in toluene for 10 min, the samples were prepared by drop casting an aliquot of 2 μl on the surface of the ATR crystals and dried them fully at open air. The analysis was conducted at an operating range of 4000 cm$^{-1}$ to 600 cm$^{-1}$ with a resolution of 4 cm$^{-1}$. On average, 128 scans were completed for each spectrum.

PL spectra were collected from diluted suspensions (100l in 1 ml of pure solvent) of nanocrystals in toluene at different times using a Horiba FluoroMax 4 spectrofluorimeter, exciting at 350 nm.

**Computational modeling.** The computational analysis was performed by designing a defect-free nanocrystal and a nanocrystal with a lattice that was displaced by half a unit cell, as is indicated in Table S2. The second structure is achieved by removing a PbBr$_2$ plane from the center of the original nanocrystal and Cs atoms from the surface in order to achieve a charge balance. Hence, we modelled the following chemical reaction:

$$\Delta E = E_{shift} - (E_{un-shifted} - N_{PbBr_2} E_{PbBr_2} - N_{CsBr} E_{CsBr})$$

where $N_{Pb}$ and $N_{Cs}$ are the number of PbBr$_2$ and CsBr molecules extracted from the full nanocrystal to get the shifted configuration. We studied three sizes, namely 2x2, 3x3 and 4x4 cubic unit cells. The energies resulted in 0.29 eV/A$^2$, 0.23 eV/A$^2$ and 0.21 eV/A$^2$, indicating that the energy needed to create the structural shift is smaller when the surface sizes are larger. All the structures were optimized under vacuum with DFT by using the PBE exchange-correlation functional[51] and a double ζ basis set plus polarization functions.[52] We accounted for scalar relativistic effects by employing effective core potential functions in the basis set. Spin–orbit coupling effects were not included in the calculations. All calculations were performed using the CP2K code.[53]All the structures were optimized under vacuum with DFT by using the PBE exchange-correlation functional[51] and a double ζ basis set plus polarization functions.[52] We accounted for scalar relativistic effects by employing effective core potential functions in the basis set. Spin–orbit coupling effects were not included in the calculations. All calculations were performed using the CP2K code.[53]

**PL characterization of heated solutions.** A vial containing 2 ml of diluted NPL solution was inserted in a home-made metal holder with holes of 4 mm diameter on the lateral sides for optical access. The holder was placed on a hot plate, and the temperature was measured with a thermocouple. PL was excited via an optical fiber coupled to a light-emitting diode emitting at 385 nm, and collected with a second fiber coupled to a spectrometer (Ocean Optics HR4000).

**Nanocrystal film preparation and PL characterization.** Films were prepared by spin-coating NPL solutions at 2000 rpm on glass substrates. For the transformation and in-situ PL measurements, the films were heated to 110 °C under air using a Peltier plate, controlled by a thermocouple sensor and a PID controller. PL was excited by a pulsed laser at 349 nm wavelength with an average power of 50 μW, and the signal was recorded with a fiber coupled spectrometer (Ocean Optics HR4000).

## Supporting Information

The Supporting Information is available free of charge on the ACS Publications website.

- ➤ Additional TEM images, pictures of vials under UV light, photoluminescence spectra, and additional data on the computational analysis.

## Corresponding Authors

*Correspondence and requests for materials should be addressed to Milena.Arciniegas@iit.it and Liberato.Manna@iit.it.

## Present Addresses

[‡]School of Engineering, Cardiff University, Queen's Buildings, The Parade, Cardiff CF24 3AA, Wales UK.

[§]Optoelectronic group. Cavendish Laboratory. University of Cambridge. J J Thomson Avenue, CB3 0HE, Cambridge, United Kingdom.

[d]IHP Leibniz-Institut für innovative Mikroelektronik, Im Technologiepark 25, D-15236 Frankfurt (Oder), Germany.

## Author Contributions

All authors contributed to the manuscript and approved the final version of it.

## Notes

The authors declare no competing financial interest.

**Acknowledgments**

Work at the Molecular Foundry was supported by the Office of Science, Office of Basic Energy Sciences, of the U.S. Department of Energy under Contract No. DE-AC02-05CH11231. ZD and MA acknowledge financial support by the EU Horizon2020 MSCA Rise project "COMPASS-691185".

**References**

1. Leijtens, T.; Bush, K. A.; Prasanna, R.; McGehee, M. D. *Nat. Energy*. **2018,** 3, (10), 828-838.
2. Stranks, S. D.; Snaith, H. J. *Nat. Nanotechnol*. **2015,** 10, (5), 391-402.
3. Ahmadi, M.; Wu, T.; Hu, B. *Adv. Mater*. **2017,** 29, (41), 1605242.
4. Droseros, N.; Longo, G.; Brauer, J. C.; Sessolo, M.; Bolink, H. J.; Banerji, N. *ACS Energy Lett*. **2018,** 3, (6), 1458-1466.
5. Protesescu, L.; Yakunin, S.; Bodnarchuk, M. I.; Krieg, F.; Caputo, R.; Hendon, C. H.; Yang, R. X.; Walsh, A.; Kovalenko, M. V. *Nano Lett*. **2015,** 15, (6), 3692-3696.
6. Huang, H.; Polavarapu, L.; Sichert, J. A.; Susha, A. S.; Urban, A. S.; Rogach, A. L. *NPG Asia Mater*. **2016,** 8, e328.
7. Sun, S. B.; Yuan, D.; Xu, Y.; Wang, A. F.; Deng, Z. T. *ACS Nano*. **2016,** 10, (3), 3648-3657.
8. Imran, M.; Caligiuri, V.; Wang, M. J.; Goldoni, L.; Prato, M.; Krahne, R.; De Trizio, L.; Manna, L. *J. Am. Chem. Soc*. **2018,** 140, (7), 2656-2664.
9. Akkerman, Q. A.; Raino, G.; Kovalenko, M. V.; Manna, L. *Nat. Mater*. **2018,** 17, (5), 394-405.
10. Shamsi, J.; Rastogi, P.; Caligiuri, V.; Abdelhady, A. L.; Spirito, D.; Manna, L.; Krahne, R. *ACS Nano*. **2017,** 11, (10), 10206-10213.
11. Shamsi, J.; Urban, A. S.; Imran, M.; De Trizio, L.; Manna, L. *Chem. Rev*. **2019**.
12. De Roo, J.; Ibanez, M.; Geiregat, P.; Nedelcu, G.; Walravens, W.; Maes, J.; Martins, J. C.; Van Driessche, I.; Koyalenko, M. V.; Hens, Z. *ACS Nano*. **2016,** 10, (2), 2071-2081.
13. Schliehe, C.; Juarez, B. H.; Pelletier, M.; Jander, S.; Greshnykh, D.; Nagel, M.; Meyer, A.; Foerster, S.; Kornowski, A.; Klinke, C.; Weller, H. *Science*. **2010,** 329, (5991), 550-553.
14. Akkerman, Q. A.; Motti, S. G.; Kandada, A. R. S.; Mosconi, E.; D'Innocenzo, V.; Bertoni, G.; Marras, S.; Kamino, B. A.; Miranda, L.; De Angelis, F.; Petrozza, A.; Prato, M.; Manna, L. *J. Am. Chem. Soc*. **2016,** 138, (3), 1010-1016.
15. Weidman, M. C.; Goodman, A. J.; Tisdale, W. A. *Chem. Mater*. **2017,** 29, (12), 5019-5030.
16. Shamsi, J.; Dang, Z. Y.; Bianchini, P.; Canale, C.; Di Stasio, F.; Brescia, R.; Prato, M.; Manna, L. *J. Am. Chem. Soc*. **2016,** 138, (23), 7240-7243.
17. Han, J. H.; Lee, S.; Cheon, J. *Chem. Soc. Rev*. **2013,** 42, (7), 2581-2591.
18. Bekenstein, Y.; Koscher, B. A.; Eaton, S. W.; Yang, P. D.; Alivisatos, A. P. *J. Am. Chem. Soc*. **2015,** 137, (51), 16008-16011.
19. Guzelturk, B.; Erdem, O.; Olutas, M.; Kelestemur, Y.; Demir, H. V. *ACS Nano*. **2014,** 8, (12), 12524-12533.
20. Cho, K. S.; Talapin, D. V.; Gaschler, W.; Murray, C. B. *J. Am. Chem. Soc*. **2005,** 127, (19), 7140-7147.
21. Van Overbeek, C.; Peters, J. L.; Van Rossum, S. A. P.; Smits, M.; Van Huis, M. A.; Vanmaekelbergh, D. *J. Phys. Chem. C*. **2018,** 122, (23), 12464-12473.
22. Du, W. M.; Qian, X. F.; Ma, X. D.; Gong, Q.; Cao, H. L.; Yin, H. *Chem.–Eur. J*. **2007,** 13, (11), 3241-3247.
23. Li, D. S.; Nielsen, M. H.; Lee, J. R. I.; Frandsen, C.; Banfield, J. F.; De Yoreo, J. J. *Science* **2012,** 336, (6084), 1014-1018.
24. Ondry, J. C.; Hauwiller, M. R.; Alivisatos, A. P. *ACS Nano*. **2018,** 12, (4), 3178-3189.
25. Nagaoka, Y.; Hills-Kimball, K.; Tan, R.; Li, R. P.; Wang, Z. W.; Chen, O. *Adv. Mater*. **2017,** 29, (18), 1606666.

26. Li, Z. J.; Hofman, E.; Davis, A. H.; Maye, M. M.; Zheng, W. W. *Chem. Mater.* **2018,** 30, (11), 3854-3860.
27. Bhaumik, S. *ChemistrySelect* **2019,** 4, (15), 4538-4543.
28. Peters, J. L.; Altantzis, T.; Lobato, I.; Jazi, M. A.; van Overbeek, C.; Bals, S.; Vanmaekelbergh, D.; Sinai, S. B. *Chem. Mater.* **2018,** 30, (14), 4831-4837.
29. Robinson, E. H.; Turo, M. J.; Macdonald, J. E. *Chem. Mater.* **2017,** 29, (9), 3854-3857.
30. Wang, Y.; Li, X. M.; Sreejith, S.; Cao, F.; Wang, Z.; Stuparu, M. C.; Zeng, H. B.; Sun, H. D. *Adv. Mater.* **2016,** 28, (48), 10637-10643.
31. Thind, A. S.; Luo, G.; Hachtel, J. A.; Morrell, M. V.; Cho, S. B.; Borisevich, A. Y.; Idrobo, J. C.; Xing, Y.; Mishra, R. *Adv. Mater.* **2018,** 31, 1805047.
32. Yu, Y.; Zhang, D. D.; Yang, P. D. *Nano Lett.* **2017,** 17, (9), 5489-5494.
33. Morrell, M. V.; He, X.; Luo, G.; Thind, A. S.; White, T. A.; Hachtel, J. A.; Borisevich, A. Y.; Idrobo, J.-C.; Mishra, R.; Xing, Y. *ACS Appl. Nano Mater.* **2018,** 1, (11), 6091-6098.
34. Akkerman, Q. A.; Bladt, E.; Petralanda, U.; Dang, Z.; Sartori, E.; Baranov, D.; Abdelhady, A. L.; Infante, I.; Bals, S.; Manna, L. *Chem. Mater.* **2019,** 31, (6), 2182-2190.
35. Di Stasio, F.; Grim, J. Q.; Lesnyak, V.; Rastogi, P.; Manna, L.; Moreels, I.; Krahne, R. *Small* **2015,** 11, (11), 1328-1334.
36. Palei, M.; Caligiuri, V.; Kudera, S.; Krahne, R. *ACS Appl. Mater. Interfaces* **2018,** 10, (26), 22356-22362.
37. Palazon, F.; Almeida, G.; Akkerman, Q. A.; De Trizio, L.; Dang, Z. Y.; Prato, M.; Manna, L. *Chem. Mater.* **2017,** 29, (10), 4167-4171.
38. Liu, Z. K.; Bekenstein, Y.; Ye, X. C.; Nguyen, S. C.; Swabeck, J.; Zhang, D. D.; Lee, S. T.; Yang, P. D.; Ma, W. L.; Alivisatos, A. P. *J. Am. Chem. Soc.* **2017,** 139, (15), 5309-5312.
39. ten Brinck, S.; Zaccaria, F.; Infante, I. *ACS Energy Lett.* **2019,** 4, (11), 2739-2747.
40. Grisorio, R.; Di Clemente, M. E.; Fanizza, E.; allegretta, i.; Altamura, D.; Striccoli, M.; Terzano, R.; Giannini, C.; Irimia-Vladu, M.; Suranna, G. P. *Nanoscale.* **2018,** 11, 986-999.
41. Yang, D. D.; Li, X. M.; Zeng, H. B. *Adv. Mater. Interf.* **2018,** 5, 170166.
42. Simons, W. W.; Laboratories, S. R., *The Sadtler Handbook of Infrared Spectra*. Sadtler: 1978.
43. Lin-Vien, D.; Colthup, N. B.; Fateley, W. G.; Grasselli, J. G., Compounds Containing –NH$_2$, –NHR, and –NR$_2$ Groups. In *The Handbook of Infrared and Raman Characteristic Frequencies of Organic Molecules*, Lin-Vien, D.; Colthup, N. B.; Fateley, W. G.; Grasselli, J. G., Eds. Academic Press: San Diego, 1991; pp 155-178.
44. Lin-Vien, D.; Colthup, N. B.; Fateley, W. G.; Grasselli, J. G., Infrared and Raman Spectra of Common Organic Compounds - Appendix 1. In *The Handbook of Infrared and Raman Characteristic Frequencies of Organic Molecules*, Academic Press: San Diego, 1991; pp 423-454.
45. Ravi, V. K.; Santra, P. K.; Joshi, N.; Chugh, J.; Singh, S. K.; Rensmo, H.; Ghosh, P.; Nag, A. *J. Phys. Chem. Lett.* **2017,** 8, (20), 4988-4994.
46. Suzuki, T.; Nishi, Y.; Fujimoto, M. *J. Am. Ceram. Soc.* **2000,** 83, (12), 3185-3195.
47. Jing, H. M.; Cheng, S.; Mi, S. B.; Lu, L.; Liu, M.; Cheng, S. D.; Jia, C. L. *ACS Appl. Mater. Interfaces* **2018,** 10, (1), 1428-1433.
48. Battle, P. D.; Green, M. A.; Laskey, N. S.; Millburn, J. E.; Murphy, L.; Rosseinsky, M. J.; Sullivan, S. P.; Vente, J. F. *Chem. Mater.* **1997,** 9, (2), 552-559.
49. Detemple, E.; Ramasse, Q. M.; Sigle, W.; Cristiani, G.; Habermeier, H. U.; Keimer, B.; van Aken, P. A. *J. Appl. Phys.* **2012,** 112, 013509.
50. Stone, G.; Ophus, C.; Birol, T.; Ciston, J.; Lee, C. H.; Wang, K.; Fennie, C. J.; Schlom, D. G.; Alem, N.; Gopalan, V. *Nat. Commun.* **2016,** 7, 12572.

51. Perdew, J. P.; Burke, K.; Ernzerhof, M. *Phys. Rev. Lett*. **1997,** 78, (7), 1396-1396.
52. VandeVondele, J.; Hutter, J. *J. Chem. Phys*. **2007,** 127, (11), 114105.
53. Hutter, J.; Iannuzzi, M.; Schiffmann, F.; VandeVondele, J. *Wiley Interdiscip. Rev. Comput. Mol. Sci.* **2014,** 4, (1), 15-25.

**Table of Content**

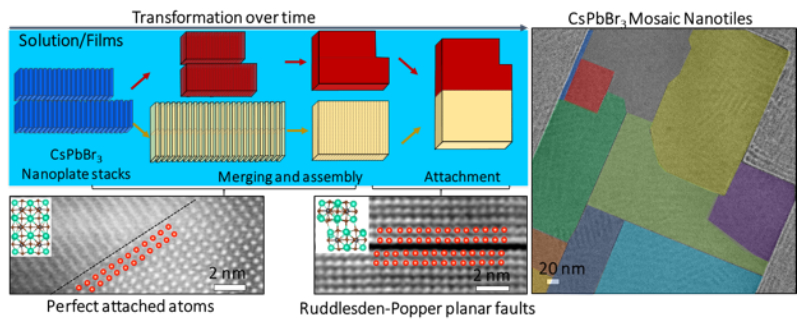